\documentclass[twocolumn,showpacs,preprintnumbers,amsmath,amssymb]{revtex4}
\usepackage{dcolumn}
\usepackage{bm,epic,eepic}
\newcommand{\beq}{\begin{equation}}
\newcommand{\enq}{\end{equation}}
\newcommand{\up}{\uparrow}\newcommand{\down}{\downarrow}
\newcommand{\ket}[1]{|#1\rangle}\newcommand{\bra}[1]{\langle#1|}

\newcommand{\av}[1]{\langle #1 \rangle}
\newcommand{\nn}{\nonumber}\newcommand{\forget}[1]{}
\newcommand{\half}{\frac{1}{2}}
\renewcommand{\vec}{\mathbf}
\renewcommand{\a}{\mathbf{a}}
\renewcommand{\b}{\mathbf{b}}
\renewcommand{\.}{\mathbf{\cdot}}
\newcommand{\pr}{^\prime}\renewcommand{\(}{\left(}
\renewcommand{\)}{\right)}

\begin{document}

\title{Quadratic Bell inequalities as tests for multipartite entanglement}

\author{Jos Uffink}
\email{uffink@phys.uu.nl}
\affiliation{%
Institute for History and Foundations of Science,\\
 Utrecht University
 PO Box 80.000, 3508 TA Utrecht, the Netherlands}%

\date{January 18, 2002; revised \today}

\begin{abstract}  This letter presents quantum mechanical inequalities
which distinguish, for systems of ${n}$ spin-$\half$ particles
(${n}>2$), between fully entangled states and states in which at
most ${n}-1$ particles are entangled. These inequalities are
stronger than those obtained by Gisin and Bechmann-Pasquinucci
[Phys.\ Lett.\ A {\bf 246}, 1 (1998)] and by Seevinck and
Svetlichny [quant-ph/0201046].
\end{abstract}
\maketitle

The Bell inequality was originally designed to test the
predictions of quantum mechanics against those of a local hidden
variables theory.
However, this inequality also provides a test to
distinguish entangled from non-entangled quantum states. Indeed,
it is well known that any non-entangled two-particle state obeys
the Bell inequalities and that all pure entangled  two-particle
states violate them for some choice of observables \cite{GISIN2}.

With the current experimental effort  to produce entangled states
of three~\cite{EXP3} and four~\cite{EXP4} particles,  it is
natural to pursue ${n}$-particle generalizations of the Bell
inequality that may likewise distinguish genuine multi-partite
entanglement from lesser entangled states.  The goal of this paper
is to report such inequalities  for spin-$\half$ particles which
are stronger than previous results \cite{SVETLICHNY,GISIN,SS}.

 The inequalities derived here are quadratic: they employ squares of
the expectation values of certain combinations of operators.
Curiously, they only provide tests for entanglement for systems of
three particles or more, and not for ${n}= 2$. At the end of this
letter, a comment is made on the question why the present
inequalities do not apply to test so-called partially separable
hidden-variables theories, as considered by
Svetlichny~\cite{SVETLICHNY} and Seevinck and
Svetlichny~\cite{SS}.

As a warming-up exercise, consider the familiar case of two
spin-$\frac{1}{2}$ particles. Let $A, A'$  denote spin observables
on the first particle, and $B, B'$ on the second. We write $AB$
etc., as shorthand for $A\otimes B$ and $\av{AB}_\rho :=
\mbox{Tr~} \rho A\otimes B$;  $\av{A B}_\psi = \bra{\psi}A \otimes
B \ket{\psi}$ for the expectations of $AB$ in the mixed state
$\rho$ or pure state $\ket{\psi}$.

The Bell inequality says that for non-entangled states, i.e., for
states of the form $\rho= \rho_1\otimes\rho_2$, or mixtures of
such states:
 \begin{equation} |\langle AB + AB' + A'B -
A'B'\rangle_\rho|  \leq 2.   \label{1} \end{equation} The maximal
violation of (\ref{1})  for  entangled
 states follows from an inequality of Cirelson~\cite{CIRELSON}
 (cf.\
 Landau~\cite{LANDAU}):
 \begin{equation} |\langle AB + AB' +
A'B - A'B'\rangle_\rho | \leq 2 \sqrt{2}. \end{equation} Equality
in (2) can be attained by the singlet state.

The first result of this paper, and the stepping stone to the
multi-particle generalizations discussed below, is that for all
states $\rho$
\begin{equation}
\av{AB' + A'B }_\rho^2 + \av{A B - A' B'}_\rho^2 \leq 4.
\label{3}\end{equation}
which strengthens the Cirelson inequality (2). (A proof  is given
in the appendix.)
 Note, however, that no smaller  bound  on the left-hand side of
(\ref{3}) exists for non-entangled states.  (To verify this, take
$\ket{\psi} = \ket{\!\!\up\up}$ and $A =A'=B=B' = \sigma_z$) Thus,
the quadratic inequality (\ref{3}) does not distinguish entangled
and non-entangled states. But we shall see below that  this is
different for multi-particle generalizations of (\ref{3}).

 Now, consider a system of three spin-$\half$ particles. In this
case, we wish to distinguish between, on the one hand, states that
are at most two-partite entangled, i.e. states of the form $\rho_1
\otimes \rho_{23}$, $\rho_2\otimes \rho_{13}$ and
$\rho_{12}\otimes \rho_{3}$, or mixtures of these states, and, on
the other hand, states which are not of this form, and are called
fully entangled. An example of a fully entangled state is the
so-called GHZ state $\frac{1}{\sqrt{2}}( \ket{\!\up\up\up} \pm
\ket{\!\down\down\down})$. Generalizations of Bell inequalities
for this purpose have been presented by Svetlichny
\cite{SVETLICHNY} and by Gisin and Bechmann-Pasquinucci
\cite{GISIN}.

As before, let $A,A'$, $B, B'$ and $C, C'$ be spin observables on
each of the three particles respectively.  Denote the set of all
three-particle states as ${\cal S}_3$ and the subset of states
which are at most two-partite entangled as ${\cal S}^2_3$.
 Svetlichny~\cite{SVETLICHNY} obtained the following
 inequalities:
 \begin{equation}
 \forall \rho \in {\cal S}^2_3:~~~~~|\av{S^\pm_3}_\rho| \leq 4,\label{S+} \end{equation}
where\forget{
\begin{eqnarray}
 {S^\pm_3} \!\!\!\! &:=& \!\!\!\!
        ABC  \mp( ABC'+ AB'C + A'B C)
 \nn\\ \!\!\!\!  &&\!\!\!\!
     - A B' C' - A' B C' - A'B'C \pm A'
 B'C'.
\end{eqnarray}}
 \begin{eqnarray}
 {S^-_3} \!\! &:=& \!\!
        ABC +ABC'+ AB'C + A'B C
 \nn\\ \!\!  &&\!\!
     - A B' C' - A' B C' - A'B'C - A'
 B'C', \\
   {S^+_3} \!\! &:=& \!\!  ABC - ABC'- AB'C - A'B C
 \nn\\    \!\!&&\!\!
  - A B'C' - A'B C' - A' B'C   + A'B'C'.
\end{eqnarray}
Ref.~\cite{SVETLICHNY} also showed that a pure state, unitarily
equivalent to the GHZ-state, yields $\av{S^\pm_3} = 4\sqrt{2}$ for
appropriate choices of observables. More recently, Seevinck and
Svetlichny \cite{SS} show  that this value is in fact the maximum
 for all three-particle states, i.e.,
\begin{equation} \forall{\rho\in {\cal S}_3}:\;\; |\av{S^\pm_3}_\rho|
\leq 4\sqrt{2}. \label{3max}
\end{equation}

Gisin and Bechmann-Pasquinucci~\cite{GISIN} obtained another
inequality by means of a recursive argument from the so-called
Bell-Klyshko inequality~\cite{BK}. Specialized to the case of
three particles, their results are:
\begin{equation} \forall \rho \in {\cal S}^2_3\;:\;\;
|\av{F_3}_\rho| \leq 2 \sqrt{2}, \label{7} \end{equation} where
\begin{equation}
F_3 :=
 ABC'+AB'C + A'B C- A' B'C'  ,\label{F3}
\end{equation} whereas
 \begin{equation}  \forall{\rho\in {\cal S}_3}\;:\;\; |\av{F_3}_\rho| \leq 4 . \label{8}\end{equation}
  Again, equality in (\ref{8}) is attained for a GHZ state and appropriate
observables.

Thus, both (\ref{S+}) and (\ref{7}) provide tests to distinguish
two-partite entangled states from fully entangled states in the
sense that a violation of either of these inequalities is a
sufficient condition for full entanglement.
 In order to compare the strength of both
inequalities,  it is useful to note that
\begin{equation} S^\pm_3 =  \mp F_3 -  F'_3, \label{comp}\end{equation}
 where $F_3'$ denotes the  same
sum of operators as $F_3$, but with all primed and unprimed
observables interchanged.  Hence, the inequalities (\ref{S+}) can
be rewritten as \begin{equation}  | \av{F_3 \pm F_3'}_\rho| \leq 4
\label{SV+}. \end{equation}
 On the other hand, since
  (\ref{7}) holds for all choices
of the observables, one can write this inequality equivalently as
 \begin{equation} \max
|\av{ F_3}_\rho| , |\av{F_3'}_\rho|  \leq 2 \sqrt{2}. \label{BKeq}
\end{equation}

It is then clear that the inequalities (\ref{SV+})  and
(\ref{BKeq}) are independent. In particular, (\ref{BKeq}) allows
the (hypothetical) case $\av{F_3} = \av{F'_3}= 2\sqrt{2}$, which
violates (\ref{SV+}), and similarly, the (equally hypothetical)
case  $\av{F_3}= 4$, $\av{F'_3}=0$ is allowed by (\ref{SV+}), but
forbidden by (\ref{BKeq}).

 \setlength{\unitlength}{0.25 mm}
 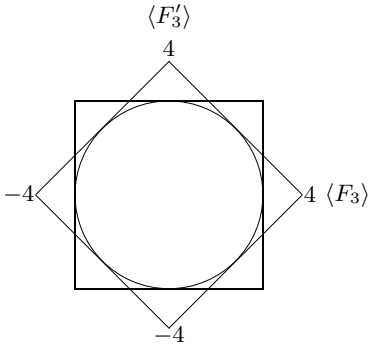
\begin{figure}[h]\begin{center}
\begin{picture}(200,180)(-100,-80)
 \put(-50,-50){\line(0,1){100}}
 \put(-50,-50){\line(1,0){100}}
  \put(50,50){\line(0,-1){100}}
 \put(50,50){\line(-1,0){100}}
\put(0,-71){\line(1,1){71}}
 \put(0,-71){\line(-1,1){71}}
  \put(0,71){\line(1,-1){71}}
 \put(0,71){\line(-1,-1){71}}
 \put(0,0){\circle{100}}
 \put(95,0){\makebox(0,0){$\av{F_3}$}}
 \put(0,95){\makebox(0,0){$\av{F'_3}$}}
 \put(0,78){\makebox(0,0){$4$}}
  \put(75,0){\makebox(0,0){$4$}}
   \put(0,-75){\makebox(0,0){$-4$}}
    \put(-80,0){\makebox(0,0){$-4$}}
\end{picture}\end{center}
\caption{Comparing the regions  in the $(\av{F_3},
\av{F'_3})$-plane allowed by the inequalities (\ref{BKeq})
(horizontal square); (\ref{SV+}) (tilted square), and (\ref{new})
(circle with radius $2\sqrt{2}$).}
\end{figure}

However, there exists a quadratic inequality that strengthens both
(\ref{SV+}) and (\ref{BKeq}). In fact,
 \begin{equation} \forall \rho \in {\cal S}^2_3 :\;\;\av{S^+_3}_\rho^2 + \av{S^-_3 }_\rho^2 \leq
 16 ,   \label{newS}
\end{equation} or equivalently, in view of  (\ref{comp}),
\begin{equation}
\forall \rho \in {\cal S}^2_3:\;\; \av{F_3}_\rho^2 +
\av{F'_3}_\rho^2 \leq 8. \label{new}
\end{equation}
\emph{Proof of (\ref{new}):} Assume, for the moment, that the
state is of the form
\begin{equation} \rho = \rho_{1 2}\otimes \rho_{3}. \label{special}\end{equation}
In that case, the expectations for particle 3  factorize from
those for the  other particles, to yield
\begin{eqnarray}\av{F_3}^2 + \av{F'_3}^2 \!\!  & =& \!\!\big(
 \av{X}\av{C} +
 \av{Y}\av{C'} \big)^2 \nonumber\\
\!\! && \!\! + \big(
 \av{X}  \av{C'}
-\av{Y}  \av{C} \big)^2 \nonumber
\\
\!\!&=&\!\!
(\av{X}^2 + \av{Y}^2)  (\av{C}^2 + \av{C'}^2) \nn\\
\!\!&\leq&\!\! 8   \label{16}
\end{eqnarray}
where I have abbreviated
 \begin{equation} X := A B'+  A'B ,~~~~
  Y := A B -
A'B'\end{equation}
 and used $\av{C}^2 + \av{C'}^2 \leq 2$,
 and  inequality (\ref{3}).
 The proof is completed by noting
 that
the left-hand side of (\ref{new}) is  invariant under a
permutation of the particle labels. Therefore, once established
for states of the  special form (\ref{special}), relation
(\ref{new}) is also true for $\rho_2 \otimes \rho_{13}$ and for
$\rho_{1}\otimes\rho_{23}$. Moreover, the left-hand side of
(\ref{new})  is a convex  function of $\rho$. Thus,  (\ref{new})
holds also for any mixture of the states just mentioned, i.e., for
all states in ${\cal S}^2_3$.

The quadratic inequality (\ref{new}) is supplemented by a  similar
weaker bound for arbitrary states:
\begin{equation} \forall{\rho \in{\cal S}_3}:\;\;
\av{F_3}_\rho^2 + \av{F'_3}_\rho^2 \leq 16
\label{newa}\end{equation}
\emph{Proof of (\ref{newa}):}
\begin{eqnarray}\sup \left(\av{F_3}^2 + \av{F'_3}^2\right)
\!\!\!\!\!\!\!\!\!\!\!\!\!\!\!\!\!\!\!\!\!\!\!\!\!\!\!\!&&\nn\\
\!\! &=& \!\! \sup \left(\av{XC + YC'}^2 + \av{ X C'- YC}^2 \right)\nn\\
\!\!&=& \!\!\sup\left( \av{XC}^2 + \av{ YC'}^2  + \av{ XC'}^2
+ \av{ YC }^2   \right. \nn\\
\!\! & &\!\! \left.
 +2 \av{XC } \av{ YC'}  - 2 \av{ XC'} \av{ YC}\right) \nn\\
 \!\!& \leq& \!\! \sup\big(\av{XC}^2 + \av{YC'}^2 +
    \av{XC'}^2 +  \av{YC}^2  \big) \nn\\
   &&  + 2\sup \big(  \av{XC}\av{YC'}  - \av{XC'}\av{YC} \big)\nn\\
&\leq&  2 \sup \big( \av{X}^2 + \av{Y}^2 \big) + 4 \sup |\av{X}\av{Y}| \nn\\
&\leq  & 4 \sup \big( \av{X}^2 + \av{Y}^2 \big) \nn\\
&=& 16, \label{proofnewa}\end{eqnarray} where the supremum is over
all $\rho \in {\cal S}_3$, and  I have used $\sup \av{XC}\leq \sup
|\av{X}| \sup|\av{C}|= \sup |\av{X}|$, etc., and $2|\av{X}\av{Y}|
\leq \av{X}^2 + \av{Y}^2$.

Equality in (\ref{newa}) is again attained for a GHZ state. This
shows that (\ref{new}), in contrast to its two-particle analogy
(\ref{3}), does distinguish between fully entangled states and
those that are at most two-partite entangled.

Let us now consider the general case of ${n}$ spin-$\half$
particles. Denote the spin observables on particle  $j$,
$j=1,\ldots, {n}$, as $A_j, A'_j$. Further, ${\cal S}_{n}$ stands
for the set of all ${n}$-particle states, and ${\cal
S}^{{n}-1}_{{n}}$ for its subset of those states which are at most
${n}-1$-partite entangled, defined similarly as ${\cal S}^2_3$.

 The inequalities of Ref.~\cite{GISIN} discussed above form part of a
recursive  chain, constructed as follows:
\begin{eqnarray}
&  F_{n} := \frac{1}{2}F_{{n}-1}( A_{n}+A_{n}')
+\frac{1}{2}F_{{n}-1}'(A_{n}-A_{n}'),& \label{GBPrecur}
\end{eqnarray} where $F_{{n}-1}'$ is the same expression as
$F_{{n}-1}$ but with all $A_j$ and $A_j'$ interchanged.  It is
then shown that
 \begin{eqnarray}
    \forall \rho \in {\cal S}^{{n}-1}_{n}: && ~~~|\av{F_{n}}_\rho| \leq 2^{{n}/2}. \label{v}
\\
 \forall \rho \in {\cal S}_{n} ~~~:&& ~~~|\av{
{F}_{n})}_\rho| \leq 2^{({n}+1)/2} \label{maxN}, \end{eqnarray}

 Recently, Ref.~\cite{SS}
has provided a generalization of the inequalities
(\ref{S+},\ref{3max}) to arbitrary ${n}$, namely
\begin{eqnarray}
\forall \rho \in {\cal S}^{{n}-1}_{{n}}:&&~~~ \av{S^\pm_{n}}_\rho
\leq 2^{{n}-1}
 \label{SSN}\\
 \forall \rho \in {\cal S}_{n}~~~: &&~~~
\av{S^\pm_{n}}_\rho \leq 2^{{n}-1}\sqrt 2 \label{SSNa}
\end{eqnarray} where
\begin{eqnarray} S^\pm_{{n}+1} :=  S^\pm_{n} A_{{n}+1}\mp  S^\mp_{{n}}A'_{{n}+1}
\;.  \label{SSrecur}\end{eqnarray} \forget{
\begin{eqnarray} S^+_{{n}+1} :=  S^+_{n} A_{{n}+1}-  S^-_{{n}}A'_{{n}+1}\;,
\nn\\
S^-_{{n}+1} :=  S^-_{n}A_{{n}+1}
 +
 S^+_{n}  A'_{{n}+1}\;.  \label{SSrecur}\end{eqnarray}
}
 In order to compare  these inequalities,
note that  the  recursive relations (\ref{GBPrecur},
\ref{SSrecur}) imply the following relations between the operators
$F_{n}$ and $S^\pm_{n}$. When ${n}$ is odd, and putting
${n}=2k+1$:
\begin{eqnarray}
S^{\pm}_{n} &=&  2^{k-1} \left( (-1)^{k(k\pm1)/2} F_{n} \mp
(-1)^{k(k\mp1)/2} F'_{n}\right) . \label{+odd}
 \end{eqnarray}
\forget{\begin{eqnarray} S^{+}_{n} &=&  2^{k-1} \left(
(-1)^{k(k+1)/2} F_{n} - (-1)^{k(k-1)/2}
F'_{n}\right) , \label{+odd}\\
S^{-}_{n} &=&  2^{k-1} \left( (-1)^{k(k-1)/2} F_{n} +
(-1)^{k(k+1)/2} F'_{n}\right), \label{-odd} \end{eqnarray}} When
${n}$ is even, writing ${n}=2k$:
\begin{eqnarray}
S^{\pm}_{n} &=&  2^{k-1} (-1)^{k(k\pm1)/2} F^{(\pm)},
\label{+even}
\forget{ \\
S^{-}_{n} &=&  2^{k-1}  (-1)^{k(k-1)/2} F^{(k-1)}, \label{-even}
}\end{eqnarray} where $F^{(+)}:= F$  and $F^{(-)} := F'$.

It appears from these relations that  the inequalities (\ref{v})
and (\ref{SSN}) are identical when  ${n}$ even, and independent
when ${n}$ is odd, as we have already seen in the special case of
${n}=3$. A similar remark holds for  (\ref{maxN}) and
(\ref{SSNa}).

Also in the case of  ${n}$ particles,  there are quadratic
inequalities which strengthen and unify the results just
mentioned. First, note that from (\ref{+odd},\ref{+even}) we
obtain the following identity:
\begin{equation}
 \av{S^+_{n}}^2  + \av{S_{n}^-}^2 =
    2^{{n}-2} \( \av{F_{n}}^2 + \av{F_{n}'}^2\).
\end{equation} Hence,  quadratic  inequalities may be expressed by
either pair of operators. In the present case, it is  convenient
to work with the pair $S^\pm$, since the recursive relation
(\ref{SSrecur}) is somewhat simpler than (\ref{GBPrecur}).

A straightforward  generalization of (\ref{proofnewa}) yields:
\begin{eqnarray}
 \sup_{\rho\in {\cal S}_{{n}}}\av{S^+_{{n}}}^2 + \av{S^-_{{n}}}^2
 \!\!\! &=&\!\!\!
      \sup_{\rho \in {\cal S}_{n}}\av{    S^+_{{n}-1} A_{n}  -  S^-_{{n}-1}}{A'}_{n} ^2
\nn\\
\!\!\!&& \!\!\!
 +   \av{ S^-_{{n}-1} A_{n}  +  S^+_{{n}-1}A'_{n}}^2  \nn\\
\!\!\!&\leq& \!\!\! 4 \sup_{\rho \in {\cal S}_{{n}-1}}
\av{S^+_{{n}-1}}^2 + \av{S^-_{{n}-1}}^2, \label{ge}\end{eqnarray}
which, by induction on (\ref{newa}), yields the following bound
for arbitrary quantum states:
\begin{equation}
\forall\rho \in {\cal S}_{n}:~~~~  \av{S^+_{n}}_\rho^2 +
\av{S^-_{n}}_\rho^2 \leq 2^{2{n}-1}. \label{genSSa}\end{equation}

Next, consider an ${n}$-particle state
 of the form $\rho_ = \rho_{1,\ldots,
{n}-1}\otimes \rho_{n}$. In  analogy with (\ref{16}), we find
\begin{widetext}
\begin{eqnarray} \av{S^+_{{n}}}^2 + \av{S^-_{{n}}}^2 &=&
      \av{   S^+_{{n}-1}A_{n}  -   S^-_{{n}-1} A'_{n}}^2
 +   \av{  S^-_{{n}-1} A_{n} +  S^+_{{n}-1} A'_{n}}^2  \nn\\
&=&
  \left(\av{ A_{{n}}} \av{ S^+_{{n}-1} }- \av{A'_{{n}}} \av{S^-_{{n}-1}}\right)^2
 + \left(\av{A_{{n}}}  \av{S^-_{{n}-1}}  + (\av{A'_{{n}}}  \av{S^+_{{n}-1}} \right)^2 \nn\\
&=& \left( \av{A_{{n}}}^2 + \av{A'_{{n}}}^2  \right)
\left( \av{S^+_{{n}-1}}^2 + \av{S^-_{{n}-1}}^2 \right)\nn\\
&\leq&  2 \sup_{\rho \in {\cal S}_{{n}-1}} \av{S^+_{{n}-1}}^2 +
\av{S^-_{{n}-1}}^2 \nn\\&\leq & 2^{2{n}-2)}.
\label{genSS}\end{eqnarray}\end{widetext}
 As before, this result is extended to
all ${n}-1$-partite entangled states by considerations of particle
label invariance and convexity. Relation (\ref{genSS}) is the
${n}$-particle generalization of  (\ref{newS}).

\emph{Concluding remarks.}---
 The
inequalities presented here provide experimentally feasible means
of testing whether multi-particle states are fully entangled, in
the sense that violation of (\ref{genSS}) is a sufficient
condition for full entanglement.
 These conditions may be useful,
since, as shown in reference \cite{SU}, some recent experiments
that claim to produce such entangled states did not exclude the
possibility of lesser entangled states. Note also that, for ${n}$
even, the test of the quadratic inequality (\ref{genSS}) requires
the same coincidence measurements for different
 spin settings as the linear inequalities (\ref{v},\ref{SSN}).
 Thus,
 the greater logical strength of the former is not paid for by an
 increase in
 experimental difficulty.

Secondly, a curious aspect of the  ${n}$-particle inequalities
presented here is that they are obtained from a basic quadratic
inequality (\ref{3}) for ${n}=2$, which itself, however, does not
distinguish between non-entangled and entangled states.

A final remark concerns the relation between testing the
entanglement of quantum states  and testing quantum mechanics
against hidden variable (HV) theories. In analogy to the local HV
theories tested by the traditional Bell inequalities,
Svetlichny~\cite{SVETLICHNY} and Svetlichny and Seevinck \cite{SS}
consider HV theories of ${n}$ particle systems with partial
separability. In such theories, not all  particles are assumed to
behave locally (separably) with respect to all others, but there
is always some subset  of particles that behave locally with
respect to the particles in another subset (where both subsets are
non-empty, of course). These authors show that the inequalities
(\ref{SV+}) and (\ref{SSN}) also characterize the predictions of
all partially separable HV theories. By contrast, the quadratic
inequalities (\ref{new},\ref{genSS}) reported here do not hold for
such theories. The reason for this is that these inequalities rely
on the validity of (\ref{3}) for any two-particle subsystem.
However, in a non-local HV model for two particles, the Cirelson
inequality, which follows from (\ref{3}), can be violated. Hence,
the inequalities derived here need not hold for such non-quantum
mechanical theories.

For example,  it is easy to construct a partially separable  HV
model for three particles: let the hidden variable $\lambda$ have
only two possible values, and let $AB= AB'= A'B = C= 1$, $A'B'= C'
=-1$ for one value of $\lambda$, and $AB= AB'= A'B = C= -1$,
$A'B'= C' =1$, for the other. Then one has $\av{S^\pm}_{\rm HV}
=4$, in accordance with (\ref{S+}), but violating (\ref{new}).

I thank George Svetlichny and Michiel Seevinck for fruitful and
stimulating discussions and a referee  for suggesting a major
simplification in the proof below.

\appendix

\section{proof of inequality (\ref{3})}
\forget{

  The expression $\av{X}^2 + \av{Y}^2 = \av{
AB' + A'B}^2 + \av{ A B -A' B'}^2$ is a convex function of $\rho$,
and so it will be sufficient to consider pure states only. Let $A
= {\vec{\sigma} \cdot \vec{a}}, A' = \vec{\sigma}\cdot \vec{a'}$,
 \forget{$B = {\vec{\sigma} \cdot \vec{b}}$,  and}, \ldots,
  $B' = \vec{\sigma}\cdot
\vec{b'}$,  where $\vec{\sigma}$ denote Pauli matrices and
$\vec{a}\ldots,\vec{b'}$ are unit vectors. We adopt a convenient
system of coordinates for the first particle and \emph{another}
for the second. In fact, for the first particle, let the $x$ axis
be oriented along $ \vec{a} + \vec{a'}$ and the $y$ axis along
$\vec{a}-\vec{a'}$. Let $\alpha$ denote the angle between
$\vec{a}$ and the $x$ axis.
 {\sc Appendix B} }

  The expression $\av{X}^2 + \av{Y}^2 = \av{
AB' + A'B}_\rho^2 + \av{ A B -A' B'}_\rho^2$ is a convex function
of $\rho$, and so it will be sufficient to consider pure states
only. Let $\rho = \ket{\psi}\bra{\psi}$. By the Schmidt
biorthogonal decomposition theorem we can write \begin{equation}
\ket{\psi} = p \ket{\phi_1} \ket{\chi_2}  - q
\ket{\phi_2}\ket{\chi_2}
\end{equation} where $p$ and $q$ are two positive numbers satisfying
$p^2 + q^2 =1$ and $\ket{\phi_i}$ and $\ket{\chi_j}$ form
orthonormal bases in the two-dimensional Hilbert spaces ${\cal
H}_1$ and ${\cal H}_2$ of the two particles respectively.

Now choose a system of coordinates  $x_1, y_1, z_1$ for the first
 particle such that $\ket{\phi_1} = \ket{\!\!\up}_1$, $\ket{\phi_2} =
 \ket{\!\!\down}_1$, in the $z_1$ direction, and a similar coordinate
 system $x_2, y_2, z_2$ for the other particle such that
  $\ket{\chi_1} = \ket{\!\!\up}_2$, $\ket{\chi_2} =
 \ket{\!\!\down}_2$, in the $z_2$ direction. Let further $A = \vec{a}
 \cdot \vec{\sigma}_1$, $B = \vec{b} \cdot \vec{\sigma}_2$, etc., where
 $\vec{\sigma}_i$ denotes the Pauli spin vector in ${\cal H}_i$.

 In these coordinates, one may write:
\begin{equation} \av{AB}_\psi=  - a_{z_1}b_{z_2}  - 2 pq (a_{x_1} b_{x_2}  +
a_{y_1}b_{y_2}) \end{equation}etc., and so, for given $\vec{a},
\vec{a'}, \vec{b} ,\vec{b'}$, the expression $\av{X}^2 + \av{Y}^2$
is a quadratic function of $2pq$. Hence it will attain its maximum
at one end of the range of $2pq$, either  $2pq=0$, or $2pq=1$.  In
the former case the state is factorizable and the inequality is
trivially satisfied. In the second case we have \begin{equation}
\av{X}^2 + \av{Y}^2
 = (\a\.\b\pr + \a\pr\.\b)^2 + (\a\.\b - \a\pr\.\b\pr)^2.
\end{equation}
Requiring this to be maximal with respect to variations in
$\a$, subject to $\a\.\a=1$, shows that $\a$ lies in the plane of
$\b$ and $\b\pr$; similarly $\a\pr$ lies in this plane.

Now let $\alpha, \beta, \gamma$, and $\delta$ denote the angles
from $\a$ to $\b$, from $\b$ to $\a\pr$,  from $\a\pr$ to $\b\pr$
and from $\b\pr$ to $\a$  respectively. Then \begin{eqnarray}
\av{X}^2 + \av{Y}^2 &=& (\cos\beta + \cos\delta)^2 + (\cos\alpha - \cos\gamma)^2 \nn\\
&=& 4\cos^2\(\frac{\beta + \delta}{2}\) \cos^2 \(\frac{\beta
-\delta}{2}\)\nn\\&& + 4\sin^2\(\frac{\alpha + \gamma}{2}\)
\sin^2\(\frac{\alpha -
\gamma}{2}\) \nn \\
&\leq& 4 \cos^2\(\frac{\beta + \delta}{2}\)
  + 4\sin^2\(\frac{\alpha + \gamma}{2}\)\nn\\
&=& 4
\end{eqnarray}
since $\alpha + \beta + \gamma + \delta = 2\pi$.

\end{document}